\documentclass[fleqn,10pt]{article}
\usepackage{subcaption}
\usepackage{bm}
\usepackage{authblk}
\usepackage{graphicx}
\usepackage{amsmath,amsfonts,amssymb}
\pdfoutput=1

\title{Multi-Season Analysis Reveals the Spatial Structure of Disease Spread}
\date{\vspace{-5ex}}

\author{Inbar Seroussi}
\author{Nir Levy}
\author{Elad Yom-Tov}
\affil{Microsoft, Herzeliya 46733, Israel}

\begin{document}
\maketitle

\begin{abstract}
Understanding the dynamics of infectious disease spread in a heterogeneous population is an important factor in designing control strategies. Here, we develop a novel tensor-driven multi-compartment version of the classic Susceptible-Infected-Recovered (SIR) model and apply it to Internet data to reveal information about the complex spatial structure of disease spread. 
The model is used to analyze state-level Google search data from the US pertaining to two viruses, Respiratory Syncytial Virus (RSV), and West Nile Virus (WNV). We fit the data with correlations of $R^2=0.70$, and $0.52$ for RSV and WNV, respectively. 
Although no prior assumptions on spatial structure are made, human movement patterns in the US explain 27-30\% of the estimated inter-state transmission rates. The transmission rates within states are correlated with known demographic indicators, such as population density and average age. 
Finally, we show that the patterns of disease load for subsequent seasons can be predicted using the model parameters estimated for previous seasons and as few as $7$ weeks of data from the current season. 
Our results are applicable to other countries and similar viruses, allowing the identification of disease spread parameters and prediction of disease load for seasonal viruses earlier in season.

\end{abstract}
\thispagestyle{empty}
\flushbottom
\maketitle

\section{Introduction}

The spread of infectious diseases through a population of susceptible individuals varies in time, space, and according to the characteristics of the susceptible individuals and the disease\cite{keeling2011modeling}. Modeling this spread can provide information on the spreading mechanism of the infection and can assist in designing strategies for control of the disease. One common method to model infections is compartmental models\cite{kermack2003contribution,hamer1906milroy}, which in their basic form assume a single population divided into 3 sub-groups of susceptible, infected, or recovered (SIR) individuals. This model describes a global spreading mechanism where each infected individual can infect any susceptible individual. Another common model describes local spreading\cite{sattenspiel2009geographic}, where infected individuals only spread infection to a limited subset of susceptible individuals.

In reality, many epidemics spread through a combination of two or more spreading mechanisms; hybrid spreading\cite{zhang2015optimizing}. Such spreading can be modeled by multiple compartments encompassing different demographic characteristics\cite{sattenspiel2009geographic}, various disease strains\cite{levy2018modeling,andreasen1997dynamics,gog2002dynamics,gupta1998chaos,crepey2015retrospective}, and cross-immunity in an age structured model\cite{castillo1989epidemiological,dawes2002onset,gog2002status}. 
However, fitting data from a given epidemic to these models remains a challenge owing to a dearth of data of sufficiently high spatial and temporal resolution. This problem is exacerbated the higher the resolution of the model, since the number of model parameters grows with the number of compartments of the model. Thus, in many cases researchers introduce simplifying assumptions in order to reduce the dimensionality of the models\cite{grenfell2001travelling,viboud2006synchrony,balcan2009multiscale}. In the absence of actual data, the spreading topology can generally be analyzed by numerical simulation of models using synthetic data, for example, of the patterns of human movements at the desired scale\cite{balcan2009multiscale,ferguson2005strategies}. It can also be analyzed through the use of theoretical tools to detect critical phenomena in networks\cite{wang2016statistical,keeling2005networks,zhang2015optimizing}. 

Another approach to overcome a lack of epidemiological data is to use proxy data, such as mobile phone data\cite{tizzoni2014use} or Internet data\cite{oren2018respiratory}. Internet data, including search engine queries, social media postings\cite{yom2015estimating,wagner2017estimating}, or Wikipedia visits
can serve as a proxy for such information and provide data at high spatial and temporal resolutions from very large populations. One demonstration of the value of these data is the recent evaluation of the childhood influenza vaccine campaign in England\cite{wagner2017estimating}, which showed that vaccination of primary school-age children significantly reduces influenza rates as estimated from Twitter posts and Bing queries. 

In this study, we focus on modeling the spatial and temporal spread of infectious disease using a tensor-driven multi-compartment version of the classic Susceptible-Infected-Recovered (SIR) model. We illustrate the ability of the model to capture disease parameters by fitting it to Internet data on two common viruses, Respiratory Syncytial Virus (RSV) and West Nile Virus (WNV) in the United States (US). Our results demonstrate that the spatial and temporal dynamics of these viruses can be captured by this model with almost no a-priori assumptions on the spreading topology and interaction. This allow us to characterize the spread of the virus of future seasons. The inferred model parameters are shown to be correlated with population indices such as human movement dynamics, and with demographic and environmental factors which are likely modulating transmission.

\section{The model}

The proposed model is a generalization of the single population SIR model\cite{kermack2003contribution}, that describes the average evolution of
an infectious disease in a single population. In the SIR model, disease spread is represented by a system of ordinary differential equations for the susceptible $S$, infected $I$, and recovered $R$ individuals. It accounts for only one pathogen and assumes equal probability of being infected by any individuals regardless of his position i.e. mean field interaction. 

The model equations are:
\begin{equation}
	\frac{dS}{dt} = - \beta S I , \quad
    \frac{dI}{dt} = \beta S I - \gamma I, \quad
    \frac{dR}{dt} = \gamma I,
    \label{scalar_sir}
\end{equation}
where $\beta > 0$ is the infection rate and $\gamma > 0$ is the recovery rate. This model, Eq. (\ref{scalar_sir}), can be generalized to account for multiple viruses and spatial effects due to groups of sub-populations and virus strains, by transforming to a multidimensional representation in which $S$, $I$, $R$, $\beta$, and $\gamma$ are  tensors (Eq. (\ref{multi_SIR})). These tensors act as time-dependent state representation of the entire system, reflecting its various attributes or dimensions. That is, each two-dimensional projection represents different sub-groups e.g. viruses, spatial regions. Here, the system describes the evolution of an infectious sub-population interacting with other sub-populations under the influence of multiple viruses.  
The dynamics for the system’s state tensors evolution is represented by the following system of ordinary differential equations: 
\begin{equation}
\label{multi_SIR}
	\begin{gathered}
    \begin{aligned}
		\frac{d\bm{S}}{dt} &= -\bm{I}\bm{\beta}^\intercal \bm{S} \\
  	  	\frac{d\bm{I}}{dt} &= \bm{S}\bm{\beta}^{\intercal}\bm{I}-\bm{\gamma}^{\intercal}\bm{I} \\
  	  	\frac{d\bm{R}}{dt} &= \bm{\gamma}^{\intercal}\bm{I},\\
	\end{aligned}
	\end{gathered}
\end{equation}
where boldface represents a tensor. We refer to this model as a multi-compartment SIR model (mcSIR). This model, and its reduction to account for multiple viruses and virus strains in one population, is presented by Levy et al.\cite{levy2018modeling}. Thus, the multi-compartment model is represented by a multi-dimensional set of equations. This representation of sub population and viruses can also be generalized to other epidemic models, such as the SIRS and the SIS\cite{keeling2011modeling}.

In this work, we focus on the spreading mechanism in a group of $N$ sub-populations (different spatial regions) analyzed separately for each virus. Within each sub-population the interaction between individual is well mixed i.e., of mean field type. In this case, $\bm{S}$, $\bm{I}$, and $\bm{R}$ take the form of square matrices with dimension $N$. The specific elements in these matrices, e.g. $S_{ij}(t)$, are defined as the fraction of susceptible individuals traveling from region $i$ to $j$ at time $t$, where  $i=j$, represents the fraction of susceptible individuals in sub-population $i$ present at time $t$. The off-diagonal elements of the matrix $\bm\beta$ account for the infection rates between sub-populations and the diagonal elements represent the infection rates within each sub-population. Note that, $\bm\beta$ is not a symmetric matrix, as the probability to move from $i$ to $j$ is not assumed to be the same as moving from $j$ to $i$. The matrix $\bm\gamma$, under the assumption of distinct sub-populations, is defined as a diagonal matrix. Each element, $\gamma_{ii}>0$, is the average recovery rate for individuals belonging to the $i$th sub-populations. For our purposes, we also assume that the number of infected people in each region is the sum of all the infected people traveling to this region, and that the number of susceptible people traveling is negligible compared to the sizes of the susceptible sub-populations in the region. In this case, the model is reduced to a traditional sub-population model\cite{sattenspiel2009geographic}.

The advantage of using a sub-population model is that in contrast to situations where populations might be well mixed (a common assumption of models without spatial structure) local disease spread is taken into account. This means that most of the population is not exposed to the infection immediately upon its introduction. Rather, the disease may have to pass through many intermediate individuals before reaching all members of the population. This is mainly due to the fact that at a larger scale, individuals that are close together in space are likely to come into contact with each other more frequently than individuals that are more distant. In addition, this locally structured populations may exhibit clique behavior. Moreover, in many cases one can find a spatial epidemic wave pattern depending on the virus type and location\cite{chattopadhyay2018conjunction,oren2018respiratory,riley2007large}. This assumption, which to some extent holds true even today when people move longer distances over shorter times, may smooth-out the clique behavior. Understanding the spatial nature of disease transmission has important consequences for control measures aimed at a disease per location. 

In order to validate the use of sub-population models presented in Eq. (\ref{multi_SIR}) to describe the observed Internet data, we tested the model on seasonal data derived from search engine queries about two commonly spread viruses in the US, RSV and WNV.
In this case, the sub-populations are the 50 states. We chose these two viruses to fit to an SIR model since there is no vaccine or medicine for prevention or treatment. Moreover the symptoms are usually mild meaning people are less likely to visit a doctor or have the case reported to the national health care authorities. On the other hand, Internet search data in these cases may provide real time high resolution estimate for the disease activity. 

\section{Methods}

\subsection{Data sources}
We used two Internet data sources in our analysis: Google Trends\cite{GoogleTrendsWN} and Twitter. Google Trends was used to determine the relative number (per state in the US) of queries made about each of the two viruses as described below.  Data was extracted at a weekly resolution. Twitter was used to estimate human mobility patterns as described below.

\subsubsection{Respiratory Syncytial Virus (RSV)} 
Data about RSV were extracted for the years 2013 to 2018 i.e., 5 seasons. A season starts at the beginning of October every year. The data comprised of the number of queries for the term ``RSV''. A single term was sufficient in this case because of the observation\cite{oren2018respiratory} that this number is in high correlation with the number of people reported as infected in each state, as published by the Centers for Disease Control and Prevention (CDC). Our data included all US states excluding Hawaii. 

\subsubsection{West Nile Virus (WNV)} 
Data about WNV were extracted for the years 2014 to 2018 i.e., 4 seasons. A season starts at the beginning of April every year. Preliminary analysis of Google Trends data showed that the number of queries for the term "West Nile" from each state during 2017 was highly correlated with the yearly total number of infected cases reported by the CDC, reaching $R^2=0.58$ (p-value=$1.4\cdot 10^{-5}$). Thus, we used the number of people querying for "West Nile" from 2014 to 2018 as a proxy for WNV disease load. Our data included all contiguous US states, excluding Hawaii and Alaska. 

\subsubsection{Data on human mobility patterns in the US}
We collected all messages from Twitter having a GPS location in the US from October 1, 2015 through April 31, 2016. For each message we extracted an anonymized user identifier, the time of the message, and the location from where it was made. These data comprised of approximately 50 million messages. The exact GPS location of each message was mapped as its encompassing US state. From these data we created a matrix of mobility patterns, where the $(i, j)$-th cell comprised of the number of people whose location in one tweet was state $i$ and in their following tweet was state $j$. This matrix was normalized by dividing the number of people moving from state $i$ to state $j$ by the total number of people who moved from state $i$ to any other state.

\subsubsection{\label{subsubsec:StateVar}Demographic information}
We extracted from the US census\cite{USCencus} variables that may affect the spreading rate of a virus within a population. These data, extracted per state, included:

\begin{enumerate}
\item Population size (number of people)
\item Density per square mile of land area.
\item Children in the following age groups: 0-4, 5-11,12-14, 14-17 (\%) 
\item Percentage of people in each of 6 main racial groups
\item Average income per household and per family (Dollars)
\item Poverty rate in adults and child under 18 (\%)
\item Senior people (65 years and over) in the population (\%)
\item Average age
\item Urban population (\%)
\item Unemployment rate (\%).
\end{enumerate}

\subsection{Parameter matching}

Fitting is carried out in two steps. In the first step, a dictionary of functions is generated. Each function is the infection rate function from an instantiation of a solution to the single population SIR model (Eq. (\ref{scalar_sir})) chosen from a wide range of parameter values. Specifically, we used $\gamma\in[10^{-4},0.1]$ with $50$ steps and $\beta\in[10^{-4},1]$ with $200$ steps. Each function entry is intended to be proportional to the number of infected individuals per state over time sampled at a weekly resolution. 
Out of the dictionary, one function is chosen, such that it has the highest correlation with the data from each state over a period of one season. 

In the second step, we fit the data to the multi-compartment SIR (mcSIR) model presented in Eq. (\ref{multi_SIR}). The parameters found in the previous step are used as an initialization to the mcSIR fitting process. Specifically, the on-diagonal elements of the matrix $\bm\beta$ were initialized to the values found from the single SIR model. The off-diagonal values of $\bm\beta$ were initialized to uniformly distributed random values between 0 and the average $\beta$ found from the fit to the single population SIR. We perform a search in the parameter space for the solution of the mcSIR model which maximizes the highest geometric mean across states of the correlations between the mcSIR estimated infection rates and data acquired from Google Trends calculated over four seasons. 

The use of multiple seasons provides us with more data to increase robustness to noise, under the assumption that the change in the infection rate from season to season and the recovery rate over the years can be accounted for by a single season-dependence multiplicative scaling factor $D$.  

Another assumption made in the fitting of the mcSIR model is the use of a scalar recovery rate $\gamma$ for all seasons and all states. This assumption was made because of the observation from the single SIR fit that the recovery rates are similar for all states and seasons. 

The gradient descent optimization process was limited to $15$ iterations. We repeated the optimization process $100$ times, each time with random initialization on the off-diagonal terms of the $\bm\beta$ matrix. The run with highest correlation between the data and the infected rate estimated by the mcSIR model was chosen.

The single population SIR and mcSIR models are solved numerically. All analyses were performed using Matlab R2017b\cite{MatlabSMLTB}. 

\section{Results}

The algorithm reached an average correlation (across states) $R^2$ of 0.70 for RSV and $R^2$ of 0.52 for WNV between the Google Trends data and the mcSIR model.

\begin{figure}
\begin{center}
	\includegraphics[trim={5cm 11cm 4cm 11cm},clip]{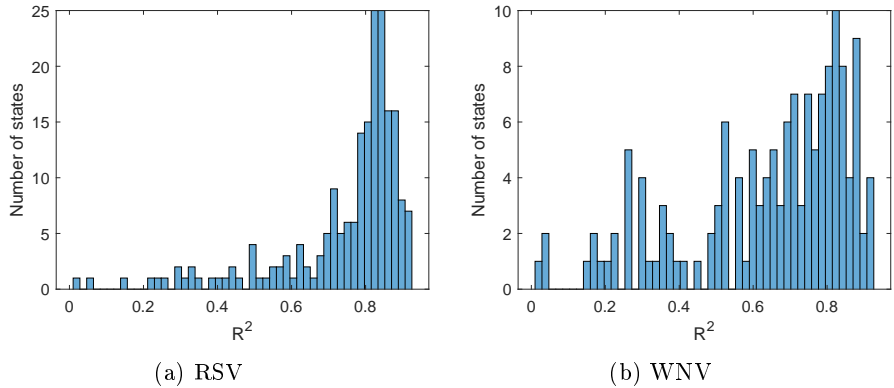}
\end{center}
\caption{\label{fig:50StateFit} An histogram of the correlation achieved between the predicted infected rate in each state and the search queries from Google Trends  (a) RSV over 4 seasons (b) WNV over 3 seasons.}
\end{figure}

Figure \ref{fig:50StateFit} shows the distribution of the model fit ($R^2$) between the Google Trends data and the mcSIR estimated infection rate in each state over $4$ seasons for RSV ($200$ state and season pairs), and $3$ seasons for WNV  ($147$ state and season pairs). For both viruses, the majority of states show high correlation between mcSIR model and the Google Trends data. For RSV (Figure \ref{fig:50StateFit}a) the correlation distribution is centered around $0.8$.  Although, the data for WNV  (Figure \ref{fig:50StateFit}b) was noisier, there was still a significant number of states with a correlation higher than $0.6$.  
We hypothesize that WNV data is noisier than RSV due to the lower incidence of the former. 

\subsection{Infection rate within states} 

After fitting the mcSIR model to the data, the infection rate matrix $\bm\beta$ and  the recovery rate $\gamma$ of each virus could be examined. We note that the best value found for the recovery rate is $\gamma=0.02$ for both RSV and WNV.  

The values along the main diagonal of the matrix $\bm\beta$ represent the infection rate within each sub-population normalized by the sub-population size\cite{sattenspiel2009geographic}. Google Trends data is normalized to the maximal number of queries in each state in a given season. This, the infection rate within each state require similar normalization. Since the  number of queries in a state are correlated with its population size, in order to estimate the infection rate in each state we multiply the diagonal elements of $\bm\beta$ by the state's population size. 

Then, in order to explain the estimated infection rate values, we built a rank regression model where the independent variables were the known demographic variables of each state (see Methods) and the estimated infection rates in each state. 

In the case of RSV, the model fit is $R^2= 0.42$ (p-value$=9.9\cdot10^{-7}$). Statistically significant variables are population density (slope: $0.64$) and average age (slope: $-0.35$). The model for WNV reached a fit of $R^2= 0.49$ (p-value$=2.92\cdot10^{-7}$). Statistically significant variables are population density (slope: $0.64$), poverty (slope: $0.31$), and average age (slope: $-0.30$). 

Infection rates of viruses increase in areas of denser population\cite{jones2008global} and so both viruses have a positive correlation with population density.  For both viruses there is a negative correlation with average age meaning that younger people are more susceptible.  RSV is known to be more severe in infants\cite{CDCRSV}. WNV is commonly spread by mosquitoes, which may explain the positive correlation with poverty, as mosquitoes could be more prevalent in such areas\cite{ladeau2013higher}. 

\subsection{Inter-state infection rates} 

The off-diagonal elements of the infection rate matrix $\bm\beta$ represent the rate of infection between states. That is, these elements represent the rate at which susceptible people in one state are likely to be infected by people from another state. Thus, we hypothesized that these elements should be correlated with human mobility patterns among states. To estimate the correlation between the state-level mobility patterns estimated from Twitter data and the values of $\bm\beta$, we normalized $\bm\beta$ by the average number of infected people estimated by the mcSIR model in each state during a particular season i.e., the infection rate matrix $\bm\beta$ is multiplied by a diagonal matrix of the average number of people transmitting the disease in a season in each state. This quantity is defined as the average transmission of the infection between different states.

The Spearman correlation between the movement patterns, as estimated from Twitter, and the estimate normalized infection rates for RSV was $\rho=0.30$ (p-value $<10^{-10}$) and for WNV was $\rho=0.27$ (p-value $<10^{-10}$). Thus, approximately a significant portion of of the variance in transmission rates of the examined viruses is explained by human mobility patterns. 

\begin{figure}
\begin{center}
  	\includegraphics[trim={2cm 1.8cm 1cm 0.7cm},clip,scale=0.55]{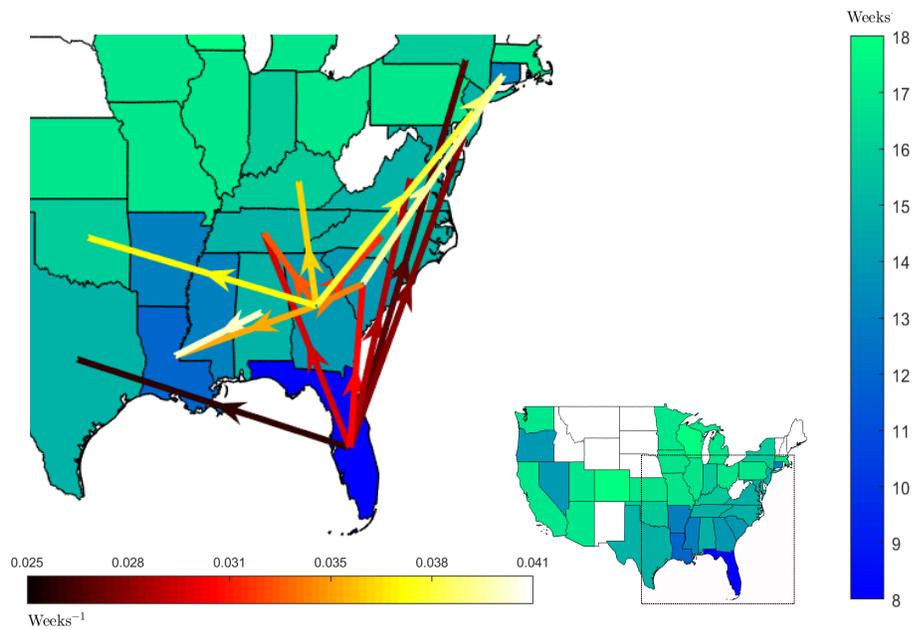}
\end{center}
\caption{\label{fig:Map2016}
Map colored according to the time to the predicted peak of the disease in each state from the beginning of the season (October $1^{st}$). The white regions represents states in which the prediction was not reliable due to insufficient and noisy data. The partial map on the left is a zoom to the region marked by the rectangular on the full map. The color of the arrows on the zoomed map signify the strength and directionality of the predicted transmission of the infection between different states of the RSV virus during the beginning of $2016$ season, as calculated from the matrix $\beta I$, averaged over the first 13 weeks in the season ($25\%$ of the season). For clarity, we show only the 16 strongest transmission rates.}
\end{figure}

Embedding the transmission rate matrix on a map of the US provides a clearer understanding of disease spread. Figure \ref{fig:Map2016} shows the transmission rate matrix averaged over the first $25\%$ of the season for the RSV virus in 2016. The color of each state represents the week of peak infection as estimated by the model. Thus, the model predicts Florida would be the first state where the disease peaks. This also identifies Florida as the source of the virus. This is in line with our knowledge of the virus spread in the US\cite{CDCRSV}. The color of the arrows in Figure \ref{fig:Map2016} represents the strength of the interaction between each two states given by the elements of the transmission matrix $\bm\beta$. Since the transmission matrix is correlated with the movements of people, the matrix elements (arrows on the map) represent the beginning of spatial virus spread in the US predicted by the model.  

\subsection{Characterization of the disease's spread in the following seasons} 

Our results suggest that the parameters of disease spread change little from year to year. This can be observed by looking at the season dependent scaling factors, $D$ (defined in the Methods). This factor approaches a value of $1$ when there is seasonal invariance. For RSV the average value of $D$, over $4$ seasons was $0.93\pm0.06$, and for WNV the average value over $3$ seasons was $0.92\pm0.08$. 
Thus, we sought to predict infection rates by applying the model with the estimated parameters from previous seasons, corrected using the first few weeks of data from the current season. This allow us to predict the infection rates of the entire season and the temporal location of the infection peak in each state. The latter is simply the extremal point of the model Eq. (\ref{multi_SIR}). 

Here, we used the estimates of $\bm\beta$ and $\gamma$ from the previews seasons, together with increasing amounts of data from the 2018 season to predict the seasonal dynamics for the entire 2018 season. Specifically, we used the parameters from the previews seasons as initial parameters for the model and adjusted them using the first $N$ weeks of data from the 2018 season, by running a few iterations of the optimization algorithm. 

\begin{figure}
\begin{center}
\includegraphics[trim={5cm 9cm 5cm 9cm},clip,scale=1]{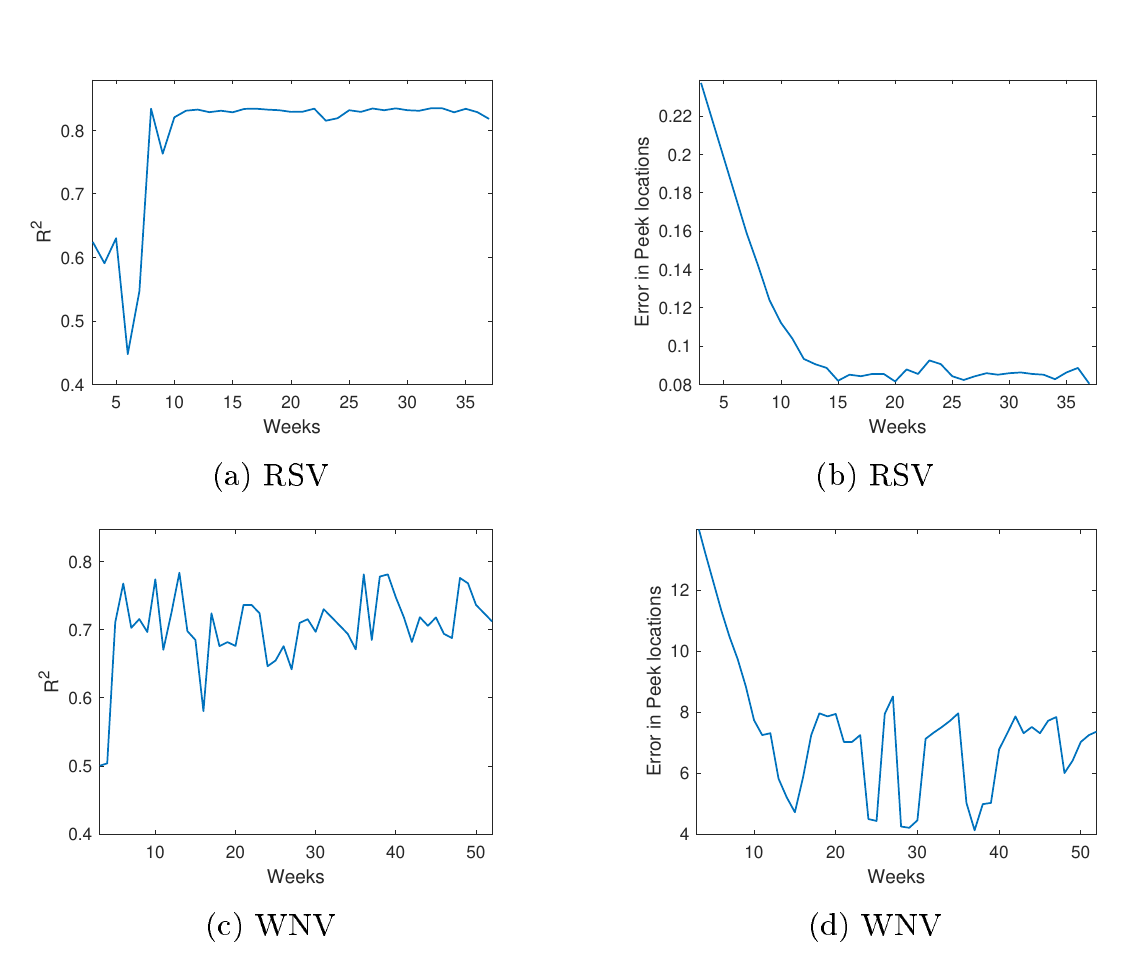}
\end{center}
\caption{\label{fig:Pred2018}(a),(c) The correlation between the percentage of infected people predicted by the model and the Google Trends data as a function of weeks, for RSV and WNV, respectively. (b), (d) The average error in the prediction of the peek of the disease for all states as a function of weeks for RSV and WNV, respectively. 
}
\end{figure}

Figures \ref{fig:Pred2018}a and \ref{fig:Pred2018}c show the correlation between the actual and predicted infection rates given increasing amounts of data for the current season (increasing $N$), for RSV and WNV, respectively. Figures \ref{fig:Pred2018}b and \ref{fig:Pred2018}d show the error in prediction of peak infection for RSV and WNV, respectively. Figure \ref{fig:Pred2018} shows that within 7 weeks from the beginning of the season it is possible to predict the progression of the entire season with reasonably high accuracy, both for RSV and WNV.

\section{Discussion}

Multi-compartment epidemic models provide a useful tool in the understanding of disease spread. However, limits on availability of epidemiological data has made the testing and validation of such models complicated. Here, we demonstrated the utility of matching proxy data from search queries to theoretical models. This matching allows us to confirm the spatial and temporal dynamics of disease spread for two sample viruses, RSV and WNV. 
Additionally, this matching provides information on the parameters of disease spread. It reveals and emphasizes the importance of accounting for the spatial structural complexity of the spreading mechanism. These parameters were shown to be correlated with the demographics of people in each state, and the mobility of people across states. 
In addition, we show that learning the parameters of the disease spread from proxy data over multiple seasons generalizes relatively well to the next season.     

Our work has drawbacks, especially, in the use of proxy data, rather than epidemiological data. Still, although less accurate, the proxy data are derived at higher spatial resolution and from a much larger, and arguably a more representative, population. The correlation between the parameters we find and the demographics of people provide an indication for the validity of this data. Therefore, this approach can be used to estimate factors influencing disease intervention and control especially in cases where epidemiological data is lacking or insufficient providing more information about the spreading mechanism.  

The modeling approach we proposed makes minimal assumptions on the structure of the population, interventions, or behavior. Our approach can be used to track and model other viruses and sub-populations. Future work will focus on modeling sub-populations (for example, different age groups) within a geographic area, as well as the effects of interventions such as vaccination.  
\bibliography{multi_SIR}

\begin{thebibliography}{10}
\expandafter\ifx\csname url\endcsname\relax
  \def\url#1{\texttt{#1}}\fi
\expandafter\ifx\csname urlprefix\endcsname\relax\def\urlprefix{URL }\fi
\providecommand{\bibinfo}[2]{#2}
\providecommand{\eprint}[2][]{\url{#2}}

\bibitem{keeling2011modeling}
\bibinfo{author}{Keeling, M.~J.} \& \bibinfo{author}{Rohani, P.}
\newblock \emph{\bibinfo{title}{Modeling infectious diseases in humans and
  animals}} (\bibinfo{publisher}{Princeton University Press},
  \bibinfo{year}{2011}).

\bibitem{kermack2003contribution}
\bibinfo{author}{Kermack, W.} \& \bibinfo{author}{Mckendrick, A.}
\newblock \bibinfo{title}{A contribution to the mathematical theory of
  epidemics}.
\newblock \emph{\bibinfo{journal}{Proceedings of the Royal Society A.}}
  \textbf{\bibinfo{volume}{115}} (\bibinfo{year}{1927}).

\bibitem{hamer1906milroy}
\bibinfo{author}{Hamer, W.~H.}
\newblock \emph{\bibinfo{title}{The Milroy lectures on epidemic disease in
  England: the evidence of variability and of persistency of type}}
  (\bibinfo{publisher}{Bedford Press}, \bibinfo{year}{1906}).

\bibitem{sattenspiel2009geographic}
\bibinfo{author}{Sattenspiel, L.}
\newblock \emph{\bibinfo{title}{The Geographic Spread of Infectious Diseases:
  Models and Applications: Models and Applications}}
  (\bibinfo{publisher}{Princeton University Press}, \bibinfo{year}{2009}).

\bibitem{zhang2015optimizing}
\bibinfo{author}{Zhang, C.}, \bibinfo{author}{Zhou, S.},
  \bibinfo{author}{Miller, J.~C.}, \bibinfo{author}{Cox, I.~J.} \&
  \bibinfo{author}{Chain, B.~M.}
\newblock \bibinfo{title}{Optimizing hybrid spreading in metapopulations}.
\newblock \emph{\bibinfo{journal}{Scientific reports}}
  \textbf{\bibinfo{volume}{5}}, \bibinfo{pages}{9924} (\bibinfo{year}{2015}).

\bibitem{levy2018modeling}
\bibinfo{author}{Levy, N.}, \bibinfo{author}{Iv, M.} \&
  \bibinfo{author}{Yom-Tov, E.}
\newblock \bibinfo{title}{Modeling {I}nfluenza-like illnesses through composite
  compartmental models}.
\newblock \emph{\bibinfo{journal}{Physica A: Statistical Mechanics and its
  Applications}} \textbf{\bibinfo{volume}{494}}, \bibinfo{pages}{288--293}
  (\bibinfo{year}{2018}).

\bibitem{andreasen1997dynamics}
\bibinfo{author}{Andreasen, V.}, \bibinfo{author}{Lin, J.} \&
  \bibinfo{author}{Levin, S.~A.}
\newblock \bibinfo{title}{The dynamics of cocirculating influenza strains
  conferring partial cross-immunity}.
\newblock \emph{\bibinfo{journal}{Journal of mathematical biology}}
  \textbf{\bibinfo{volume}{35}}, \bibinfo{pages}{825--842}
  (\bibinfo{year}{1997}).

\bibitem{gog2002dynamics}
\bibinfo{author}{Gog, J.~R.} \& \bibinfo{author}{Grenfell, B.~T.}
\newblock \bibinfo{title}{Dynamics and selection of many-strain pathogens}.
\newblock \emph{\bibinfo{journal}{Proceedings of the National Academy of
  Sciences}} \textbf{\bibinfo{volume}{99}}, \bibinfo{pages}{17209--17214}
  (\bibinfo{year}{2002}).

\bibitem{gupta1998chaos}
\bibinfo{author}{Gupta, S.}, \bibinfo{author}{Ferguson, N.} \&
  \bibinfo{author}{Anderson, R.}
\newblock \bibinfo{title}{Chaos, persistence, and evolution of strain structure
  in antigenically diverse infectious agents}.
\newblock \emph{\bibinfo{journal}{Science}} \textbf{\bibinfo{volume}{280}},
  \bibinfo{pages}{912--915} (\bibinfo{year}{1998}).

\bibitem{crepey2015retrospective}
\bibinfo{author}{Cr{\'e}pey, P.}, \bibinfo{author}{de~Boer, P.~T.},
  \bibinfo{author}{Postma, M.~J.} \& \bibinfo{author}{Pitman, R.}
\newblock \bibinfo{title}{Retrospective public health impact of a quadrivalent
  influenza vaccine in the {U}nited {S}tates}.
\newblock \emph{\bibinfo{journal}{Influenza and other respiratory viruses}}
  \textbf{\bibinfo{volume}{9}}, \bibinfo{pages}{39--46} (\bibinfo{year}{2015}).

\bibitem{castillo1989epidemiological}
\bibinfo{author}{Castillo-Chavez, C.}, \bibinfo{author}{Hethcote, H.~W.},
  \bibinfo{author}{Andreasen, V.}, \bibinfo{author}{Levin, S.~A.} \&
  \bibinfo{author}{Liu, W.~M.}
\newblock \bibinfo{title}{Epidemiological models with age structure,
  proportionate mixing, and cross-immunity}.
\newblock \emph{\bibinfo{journal}{Journal of mathematical biology}}
  \textbf{\bibinfo{volume}{27}}, \bibinfo{pages}{233--258}
  (\bibinfo{year}{1989}).

\bibitem{dawes2002onset}
\bibinfo{author}{Dawes, J.} \& \bibinfo{author}{Gog, J.}
\newblock \bibinfo{title}{The onset of oscillatory dynamics in models of
  multiple disease strains}.
\newblock \emph{\bibinfo{journal}{Journal of mathematical biology}}
  \textbf{\bibinfo{volume}{45}}, \bibinfo{pages}{471--510}
  (\bibinfo{year}{2002}).

\bibitem{gog2002status}
\bibinfo{author}{Gog, J.} \& \bibinfo{author}{Swinton, J.}
\newblock \bibinfo{title}{A status-based approach to multiple strain dynamics}.
\newblock \emph{\bibinfo{journal}{Journal of mathematical biology}}
  \textbf{\bibinfo{volume}{44}}, \bibinfo{pages}{169--184}
  (\bibinfo{year}{2002}).

\bibitem{grenfell2001travelling}
\bibinfo{author}{Grenfell, B.}, \bibinfo{author}{Bj{\o}rnstad, O.} \&
  \bibinfo{author}{Kappey, J.}
\newblock \bibinfo{title}{Travelling waves and spatial hierarchies in measles
  epidemics}.
\newblock \emph{\bibinfo{journal}{Nature}} \textbf{\bibinfo{volume}{414}},
  \bibinfo{pages}{716} (\bibinfo{year}{2001}).

\bibitem{viboud2006synchrony}
\bibinfo{author}{Viboud, C.} \emph{et~al.}
\newblock \bibinfo{title}{Synchrony, waves, and spatial hierarchies in the
  spread of {I}nfluenza}.
\newblock \emph{\bibinfo{journal}{science}} \textbf{\bibinfo{volume}{312}},
  \bibinfo{pages}{447--451} (\bibinfo{year}{2006}).

\bibitem{balcan2009multiscale}
\bibinfo{author}{Balcan, D.} \emph{et~al.}
\newblock \bibinfo{title}{Multiscale mobility networks and the spatial
  spreading of infectious diseases}.
\newblock \emph{\bibinfo{journal}{Proceedings of the National Academy of
  Sciences}} \bibinfo{pages}{pnas--0906910106} (\bibinfo{year}{2009}).

\bibitem{ferguson2005strategies}
\bibinfo{author}{Ferguson, N.~M.} \emph{et~al.}
\newblock \bibinfo{title}{Strategies for containing an emerging influenza
  pandemic in southeast asia}.
\newblock \emph{\bibinfo{journal}{Nature}} \textbf{\bibinfo{volume}{437}},
  \bibinfo{pages}{209} (\bibinfo{year}{2005}).

\bibitem{wang2016statistical}
\bibinfo{author}{Wang, Z.} \emph{et~al.}
\newblock \bibinfo{title}{Statistical physics of vaccination}.
\newblock \emph{\bibinfo{journal}{Physics Reports}}
  \textbf{\bibinfo{volume}{664}}, \bibinfo{pages}{1--113}
  (\bibinfo{year}{2016}).

\bibitem{keeling2005networks}
\bibinfo{author}{Keeling, M.~J.} \& \bibinfo{author}{Eames, K.~T.}
\newblock \bibinfo{title}{Networks and epidemic models}.
\newblock \emph{\bibinfo{journal}{Journal of the Royal Society Interface}}
  \textbf{\bibinfo{volume}{2}}, \bibinfo{pages}{295--307}
  (\bibinfo{year}{2005}).

\bibitem{tizzoni2014use}
\bibinfo{author}{Tizzoni, M.} \emph{et~al.}
\newblock \bibinfo{title}{On the use of human mobility proxies for modeling
  epidemics}.
\newblock \emph{\bibinfo{journal}{PLoS computational biology}}
  \textbf{\bibinfo{volume}{10}}, \bibinfo{pages}{e1003716}
  (\bibinfo{year}{2014}).

\bibitem{oren2018respiratory}
\bibinfo{author}{Oren, E.}, \bibinfo{author}{Frere, J.},
  \bibinfo{author}{Yom-Tov, E.} \& \bibinfo{author}{Yom-Tov, E.}
\newblock \bibinfo{title}{Respiratory syncytial virus tracking using internet
  search engine data}.
\newblock \emph{\bibinfo{journal}{BMC public health}}
  \textbf{\bibinfo{volume}{18}}, \bibinfo{pages}{445} (\bibinfo{year}{2018}).

\bibitem{yom2015estimating}
\bibinfo{author}{Yom-Tov, E.}, \bibinfo{author}{Johansson-Cox, I.},
  \bibinfo{author}{Lampos, V.} \& \bibinfo{author}{Hayward, A.~C.}
\newblock \bibinfo{title}{Estimating the secondary attack rate and serial
  interval of influenza-like illnesses using social media}.
\newblock \emph{\bibinfo{journal}{Influenza and other respiratory viruses}}
  \textbf{\bibinfo{volume}{9}}, \bibinfo{pages}{191--199}
  (\bibinfo{year}{2015}).

\bibitem{wagner2017estimating}
\bibinfo{author}{Wagner, M.}, \bibinfo{author}{Lampos, V.},
  \bibinfo{author}{Yom-Tov, E.}, \bibinfo{author}{Pebody, R.} \&
  \bibinfo{author}{Cox, I.~J.}
\newblock \bibinfo{title}{Estimating the population impact of a new pediatric
  {I}nfluenza vaccination program in {E}ngland using social media content}.
\newblock \emph{\bibinfo{journal}{Journal of medical Internet research}}
  \textbf{\bibinfo{volume}{19}}, \bibinfo{pages}{e416} (\bibinfo{year}{2017}).

\bibitem{chattopadhyay2018conjunction}
\bibinfo{author}{Chattopadhyay, I.}, \bibinfo{author}{Kiciman, E.},
  \bibinfo{author}{Elliott, J.~W.}, \bibinfo{author}{Shaman, J.~L.} \&
  \bibinfo{author}{Rzhetsky, A.}
\newblock \bibinfo{title}{Conjunction of factors triggering waves of seasonal
  {I}nfluenza}.
\newblock \emph{\bibinfo{journal}{eLife}} \textbf{\bibinfo{volume}{7}}
  (\bibinfo{year}{2018}).

\bibitem{riley2007large}
\bibinfo{author}{Riley, S.}
\newblock \bibinfo{title}{Large-scale spatial-transmission models of infectious
  disease}.
\newblock \emph{\bibinfo{journal}{Science}} \textbf{\bibinfo{volume}{316}},
  \bibinfo{pages}{1298--1301} (\bibinfo{year}{2007}).

\bibitem{GoogleTrendsWN}
\bibinfo{title}{Google {T}rends. 2017, available from:
  https://trends.google.com/trends/}.

\bibitem{USCencus}
\bibinfo{title}{United {S}tates {C}ensus {B}ureau, available from
  https://www.census.gov}.

\bibitem{MatlabSMLTB}
\bibinfo{title}{Matlab statistics and machine learning toolbox}
  (\bibinfo{year}{2017b}).
\newblock \bibinfo{note}{The MathWorks, Natick, MA, USA}.

\bibitem{jones2008global}
\bibinfo{author}{Jones, K.~E.} \emph{et~al.}
\newblock \bibinfo{title}{Global trends in emerging infectious diseases}.
\newblock \emph{\bibinfo{journal}{Nature}} \textbf{\bibinfo{volume}{451}},
  \bibinfo{pages}{990} (\bibinfo{year}{2008}).

\bibitem{CDCRSV}
\bibinfo{title}{Centers for {D}isease {C}ontrol and {P}revention, available
  from: https://www.cdc.gov/rsv/about/index.html}.

\bibitem{ladeau2013higher}
\bibinfo{author}{LaDeau, S.~L.}, \bibinfo{author}{Leisnham, P.~T.},
  \bibinfo{author}{Biehler, D.} \& \bibinfo{author}{Bodner, D.}
\newblock \bibinfo{title}{Higher mosquito production in low-income
  neighborhoods of baltimore and {W}ashington, {DC}: understanding ecological
  drivers and mosquito-borne disease risk in temperate cities}.
\newblock \emph{\bibinfo{journal}{International journal of environmental
  research and public health}} \textbf{\bibinfo{volume}{10}},
  \bibinfo{pages}{1505--1526} (\bibinfo{year}{2013}).

\end{thebibliography}
\end{document}